\documentclass[twocolumn]{aastex631}

\usepackage{amsmath}
\usepackage{changepage}

\begin{document}

\title{Tidal synchronization trapping in stars and planets with convective envelopes}

%% you desire. In v6.31 you can include a footnote in the title.

%% A significant change from earlier AASTEX versions is in the structure for 
%% calling author and affiliations. The change was necessary to implement 
%% auto-indexing of affiliations which prior was a manual process that could 
%% easily be tedious in large author manuscripts.
%%
%% The \author command is the same as before except it now takes an optional
%% argument which is the 16 digit ORCID. The syntax is:
\author[0000-0001-9420-5194]{Janosz W. Dewberry}
\affiliation{Canadian Institute for Theoretical Astrophysics, 60 St. George Street, Toronto, ON M5S 3H8, Canada}
\begin{abstract}
Tidal torques can alter the spins of tidally interacting stars and planets, usually over shorter timescales than the tidal damping of orbital separations or eccentricities. Simple tidal models predict that, in eccentric binary or planetary systems, rotation periods will evolve toward a “pseudosynchronous” ratio with the orbital period. However, this prediction does not account for “inertial” waves that are present in stars or gaseous planets with (i) convective envelopes and (ii) even very slow rotation. We demonstrate that tidal driving of inertial oscillations in eccentric systems generically produces a network of stable “synchronization traps” at ratios of orbital to rotation period that are simple to predict but can deviate significantly from pseudosynchronization. The mechanism underlying spin synchronization trapping is similar to tidal resonance locking, involving a balance between torques that is maintained automatically by the scaling of inertial mode frequencies with the rotation rate. In contrast with many resonance locking scenarios, however, the torque balance required for synchronization trapping need not drive mode amplitudes to nonlinearity. Synchronization traps may provide an explanation for low-mass stars and hot Jupiters with observed rotation rates that deviate from pseudosynchronous or synchronous expectations.
\end{abstract}



\section{Introduction}\label{sec:intro} 
Tidal torques typically affect the spins of tidally interacting bodies more rapidly than their orbital motion, because the orbital angular momentum usually dwarfs constituent spin angular momenta. Unravelling the effects of tides on spin evolution therefore stands as a prerequisite to predicting their impact on orbital semimajor axes, eccentricities, or spin-orbit misalignments. Simple tidal models invoking a constant time lag in the tidally perturbed body's response to every component of a Fourier-decomposed tidal potential predict that in eccentric systems, tidally interacting planets or stars will evolve toward ``pseudosynchronous'' states in which the ratio of the spin frequency $\Omega$ to the orbital mean motion $\Omega_o$ approaches a value that depends only on the eccentricity \citep{Hut1981}. Pseudosynchronization or synchronization of orbital and rotation periods are therefore commonly assumed as initial conditions in both theoretical and observational studies of eccentricity or obliquity damping by tides \citep[e.g.,][]{Goodman1998,Ogilvie2007,Leconte2010,Vick2019,Barker2022,Rice2022}. However, both numerical calculations \citep{Townsend2023} and some observations \citep{deWit2016,Lurie2017} suggest that this assumption may not always be warranted.

We demonstrate that, in stars or gaseous planets with large convective regions, dynamical driving of oscillations intrinsically linked to rotation (in particular, inertial modes) by eccentricity tides can hinder tidal evolution toward pseudosynchronization. Sufficiently large eccentricity aliases inertial mode resonances to tidal frequencies at which they can oppose and balance the torque from nonresonantly driven fundamental oscillations (the ``equilibrium'' tide), introducing a network of spin frequencies $\Omega$ at which the time derivative $\dot{\Omega}$ vanishes. Because inertial modes have frequencies that are directly proportional to the rotation rate of a star or planet, a subset of these spin equilibria constitute stable fixed points in the parameter $\Omega/\Omega_o$, attracting rotation rates toward particular ratios with the mean motion. We predict that many lower-mass stars and/or hot jupiters may be caught in these synchronization traps until eccentricities are damped away, with their rotation rates held at values that deviate significantly from pseudosynchronization. 

This paper is structured as follows: Section \ref{sec:mech} describes the basic mechanism underlying synchronization trapping, and Section \ref{sec:num} presents a quantitative demonstration of its efficacy. Section \ref{sec:disc} then provides a discussion of astrophysical relevance and a comparison with observations, while Section \ref{sec:sum} concludes. The appendices provide further details related to the derivation of the secular tidal evolution equations and to the numerical calculations described in Section \ref{sec:num}.

\section{Synchronization mechanism}\label{sec:mech}
Consider a rotating star with mass $M$, equatorial radius $R$, and equilibrium density $\rho_0=\rho_0(r,\theta)$, which is  tidally perturbed by a secondary mass $M_2$. We assume that the orbit of the two bodies is aligned with the spin of the primary tidally perturbed star, that the secondary is distant enough to be treated as a point mass, and that the tidal potential produced by this point mass can be approximated as purely quadrupolar. Under the additional assumption that the primary rotates rigidly, \autoref{app:teveqn} reviews the derivation of secular equations for the time evolution of the rotation rate $\Omega$, the orbital semimajor axis $a,$ and the orbital eccentricity $e.$ The spin evolution is governed by
\begin{align}\label{eq:Odot}
    \frac{\dot{\Omega}}{\Omega}
    &=\frac{3}{2}\frac{GMM_2}{aI\Omega}\frac{M_2}{M}
    \left(\frac{R}{a}\right)^5
    \sum_{k=-\infty}^\infty
    |X_{22k}|^2\kappa_{22k},
\end{align}
while the eccentricity and semimajor axis evolve according to \autoref{eq:edot} and \autoref{eq:adot}, respectively. Here, $I=\int_V\rho_0R^2\text{d}V$ is the primary's moment of inertia, $\Omega_o=[G(M+M_2)/a^3]^{1/2}$ is the mean motion, $X_{\ell mk}$ are Hansen coefficients (\autoref{eq:hansen}), and $\kappa_{\ell mk}=\Im[k_{\ell m}]$ are the imaginary parts of tidal Love numbers $k_{\ell m}=k_{\ell m}(k\Omega_o)$ describing the response of the star in spherical harmonic degree $\ell$, azimuthal wavenumber $m,$ and tidal frequency $k\Omega_o$ to a tidal potential with the form $\Re[A_{\ell mk}(r/R)^\ell Y_{\ell m}e^{-\text{i}k\Omega_ot}]$ \citep[see Appendix \ref{app:pot}; ][]{Ogilvie2014}.

Spin evolution takes place on a much shorter timescale than semimajor axis or eccentricity evolution, because the orbital angular momentum of the binary dwarfs the spin angular momentum of the primary. Quantitatively, the right-hand sides of Equations \eqref{eq:Odot}, \eqref{eq:adot}, and \eqref{eq:edot} involve similar coefficients, but with prefactors in the dimensionless ratio
\begin{equation}\label{eq:scale}
    \frac{GMM_2}{ka\Omega_oI\Omega}
    \approx 
    \frac{1.4\times 10^4 q}{k(1+q)^{1/3}}
    \left(\frac{M}{M_\odot}\right)^{\frac{5}{3}}
    \left(\frac{J_\odot}{J}\right)
    \left(\frac{P_o}{10\text{d}}\right)^{\frac{1}{3}}
\end{equation}
where $q=M_2/M$ is the mass ratio, $J=\int_V\rho_0R^2\Omega\text{d}V$ is the spin angular momentum, and $P_o=2\pi/\Omega_o$ is the orbital period. For this reason, studies of tidal evolution in (for example) solar-mass binary star systems commonly assume that spin synchronization can be taken as a foregone conclusion prior to considering the effects of tides on $a$ and $e$ \citep[e.g.,][]{Goodman1998,Barker2022}. We do not necessarily disagree. However, self-consistently including the effects of stellar rotation on the dissipative coefficients $\kappa_{\ell mk}$ can lead to equilibrium synchronization states (with $\dot{\Omega}=0$) that deviate significantly from the pseudosynchronous rotation rate predicted by \cite{Hut1981}.

\subsection{Synchronization equilibria}\label{sec:eqm}
To search for equilibrium rotation states with $\dot{\Omega}=0$, first note that 
as $e\rightarrow0,$ $X_{222}\rightarrow1,$ $X_{221}\rightarrow-e/2,$ and $X_{223}\rightarrow7e/2$, while $X_{22k}\sim\mathcal{O}(e^2)$ for all other $k.$ Then to second order in eccentricity,
\begin{equation}\label{eq:Odote2}
    \frac{\dot{\Omega}}{\Omega}
    \approx\frac{3}{2}\frac{GM_2^2}{aI\Omega}
    \left(\frac{R}{a}\right)^5
    \left[ 
        \kappa_{222} 
        +\frac{e^2}{4}\left(
            \kappa_{221} 
            +49\kappa_{223} 
        \right)
    \right].
\end{equation}
Additionally, positive definite dissipation requires that $\text{sgn}[\kappa_{\ell mk}]=\text{sgn}[k\Omega_o-m\Omega]$ \citep{Ogilvie2013}. Therefore when $e=0$ (in which case $X_{\ell m k}=0$ unless $m=k$), $\dot{\Omega}$ can vanish only  when $\Omega_o=\Omega$ (i.e., when the spin synchronizes with the orbit). Nonzero eccentricity alters the synchronization state; if all of the dissipative coefficients conveniently satisfy the linear relation $\kappa_{\ell mk}\propto(k\Omega_o - m\Omega)$ with the same constant of proportionality, $\dot{\Omega}$ vanishes only at the pseudosynchronous ratio of $\Omega/\Omega_o$ predicted by \citet{Hut1981}: 
\begin{equation}\label{eq:hut}
    \frac{\Omega }{\Omega_o}
    =\frac{1 + (15/2)e^2 + (45/8)e^4 + (5/16)e^6}
    {[1 + 3e^2 + (3/8)e^4](1 - e^2)^{3/2}}.
\end{equation}

\begin{figure*}
    \centering
    \includegraphics[width=\textwidth]{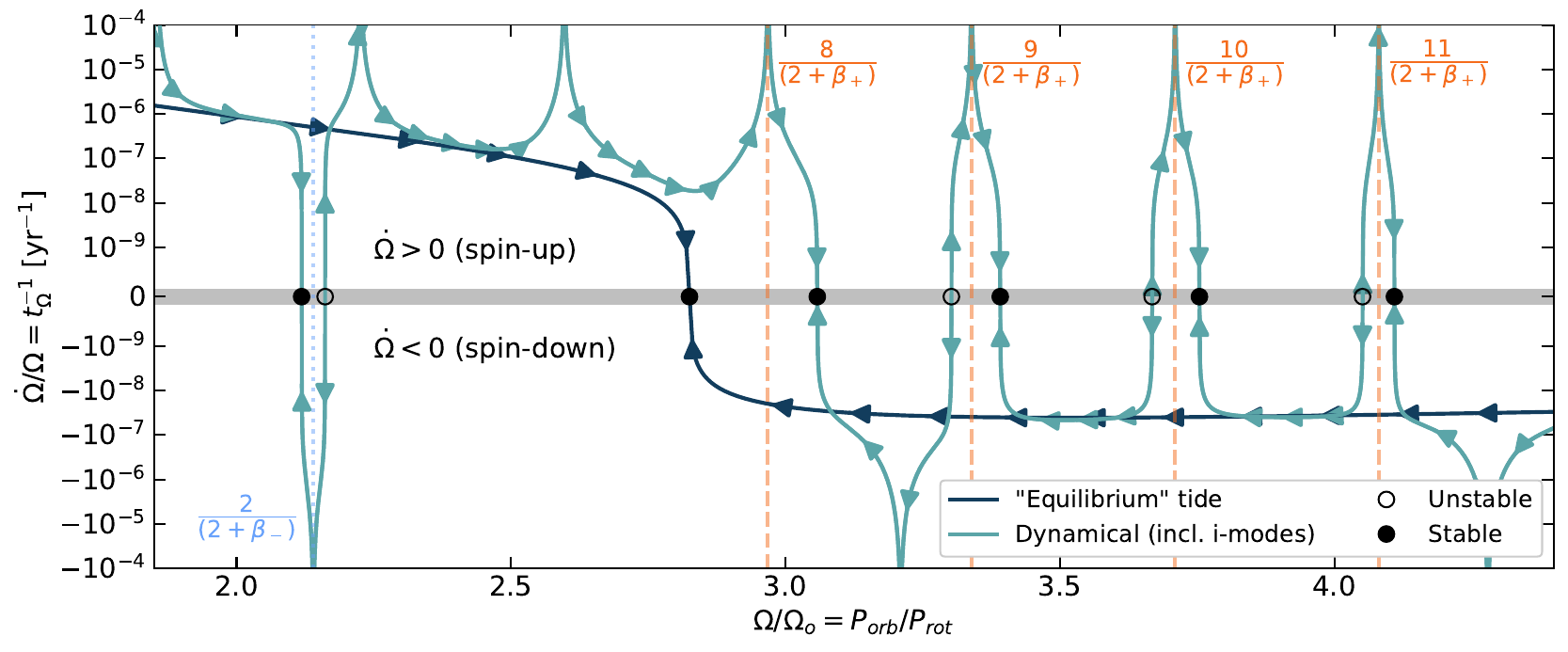}
    \caption{{Profiles of $\dot{\Omega}/\Omega$ (right-hand side of \autoref{eq:Odot}) computed for an $n=1.5$ model of a fully convective, rotating star (with rotation rate $\Omega\simeq0.09\Omega_d,$ where $\Omega_d$ is the star's dynamical frequency), which is tidally interacting with an equal mass body through an orbit with eccentricity $e=0.5$, and a varying orbital mean motion $\Omega_o$ (the y-axis switches from log to linear scale at $10^{-9}$). We adopt a constant kinematic viscosity $\nu$ with Ekman number $E_k=\nu/(R^2\Omega)=10^{-5}$. The dark blue solid line plots a profile computed solely with f-modes, and roughly describes the ``equilibrium tide'' of weak frictional theories \citep{Ogilvie2014}. The light blue line plots a profile of $\dot{\Omega}/\Omega$ computed with the inclusion of inertial modes, which introduce additional rotational equilibria (where $\dot{\Omega}=0$). The arrows along the curves indicate the expected evolution toward (away from) the stable (unstable) fixed points indicated by the filled (unfilled) black points. }}
    \label{fig:demo}
\end{figure*}

Strong nonmonotonic variation in the $\kappa_{\ell mk}$ can introduce additional rotation rates at which $\dot{\Omega}=0$, though, specifically when tidal frequencies come into resonance with the natural frequencies of the seismic oscillation modes of the tidally perturbed body. Each $\kappa_{\ell mk}$ can be computed from the properties of oscillations via \citep[e.g.,][]{Dewberry2024}
\begin{equation}\label{eq:kapp}
    \kappa_{22k}
    =\frac{2\pi}{5}
    \sum_{\alpha}
    \frac{|Q_{22}^\alpha|^2\gamma_k}
    {\epsilon_{\alpha}(\Delta_{\alpha k}^2 +\gamma_k^2)},
\end{equation}
where $\alpha$ labels modes, and $\Delta_{\alpha k}=\omega_\alpha-\omega_k$ is the detuning between the tidal frequency $\omega_k=k\Omega_o-m\Omega$ and the frequency $\omega_\alpha$ of a given mode $\alpha$ (both considered in the rotating frame). Meanwhile $\epsilon_\alpha$ and $Q_{\ell m}^\alpha$ are constants characterizing the mode energy and spatial overlap with the tidal force (see \autoref{app:mode}). Lastly, $\gamma_k=\gamma(k\Omega_o)$ is the (tidal frequency dependent) damping rate experienced by all of the collectively driven oscillations due to internal processes.\footnote{
This $\gamma_k$ agrees with individual mode damping rates $\gamma_\alpha$ when the detuning $\Delta_{\alpha k}\rightarrow0$, but can deviate at other frequencies \citep{Townsend2023,Dewberry2024}.
} We refer to all quantities appearing in \autoref{eq:kapp} in a dimensionless form scaled by units with $G=M=R=1.$

Away from resonances, \autoref{eq:kapp} is dominated by monotonic contributions from nonresonantly driven ``fundamental'' (f-)modes. In the absence of any other oscillation modes, \emph{nonresonant} f-mode driving predicts a single pseudosynchronous state in rough agreement with constant time-lag models (see the dark blue solid line in Fig. \ref{fig:demo}). On the other hand, small detunings $\Delta_{\alpha k}$ (combined with weak damping and large ratios $|Q_{\ell m}^\alpha|^2/\epsilon_\alpha$) can introduce additional zero-crossings for $\dot{\Omega}$ curves computed for a given set of orbital parameters, when the nonresonant f-mode torque is balanced by a near-resonant torque contribution of the opposite sign (see the light blue solid line in Fig. \ref{fig:demo}). 

If the dependence of the mode's rotating frame frequency $\omega_\alpha$ on $\Omega$ is known, the resonance condition $\omega_\alpha - \omega_k\simeq0$ in principle predicts these additional equilibrium rotation states. The rotating-frame frequencies of, for example, gravito-inertial oscillations can be written as $\omega_\alpha\simeq \omega_{\alpha0} + \beta_\alpha\Omega$, where $\omega_{\alpha0}$ is the mode's frequency in the nonrotating limit and $\beta_\alpha$ is a constant that depends on internal structure. Balance with $\omega_k$ would then imply the existence of equilibrium rotation states close to the ratio
\begin{equation}\label{eq:gres}
    \frac{\Omega}{\Omega_o}
    \approx\frac{k - \omega_{\alpha0}/\Omega_o}
    {m + \beta_\alpha}.
\end{equation}
\autoref{eq:gres} is less useful than \autoref{eq:hut}, though, since it predicts a rotation rate $\Omega$ that does not scale directly with the mean motion $\Omega_o;$ as the orbit changes, so too will the ratio $\Omega/\Omega_o$. Both $\omega_{\alpha0}$ and the coefficient $\beta_\alpha$ are also subject to change with the internal structure of the tidally perturbed body as it evolves. 

Several previous works have noted that eccentricity can introduce multiple possible spin equilibria \citep[e.g.,][]{Storch2014}. The main point of this paper is to highlight the fact that the dynamics simplify when oscillations possess frequencies that vanish in the limit of zero rotation ($\omega_{\alpha0}=0$). In particular, planets and stars with convective envelopes possess long-wavelength inertial modes \citep[e.g.,][]{Wu2005} or mode-like flows \citep{Papaloizou2010,Lin2021} that enhance dissipation at frequencies that are (at least in the limit of slow-to-moderate rotation) linearly related to $\Omega$; inertial mode frequencies satisfy $\omega_\alpha\approx\beta_\alpha\Omega,$ where $\beta_\alpha$ depends only weakly on internal structure \citep[e.g.,][]{Dewberry2022}. We find that, in stellar models with a large convective envelope and for eccentricities $\gtrsim0.1$, resonances with the pair of prograde and retrograde $m=2$ inertial modes with the longest wavelengths (for which we write $\beta_\alpha=\beta_{\pm}$) introduce numerous stable synchronization states close to the ratios
\begin{equation}\label{eq:res}
    \frac{\Omega}{\Omega_o}
    \approx\frac{k}{2+\beta_\pm}.
\end{equation}

The stable subset of these equilibria are indicated by the black filled points plotted at some of the light blue curve's zero-crossings in \autoref{fig:demo}. The key point is that, because inertial-wave frequencies scale directly with $\Omega$, these synchronization states remain fixed near the ratio of $\Omega/\Omega_o$ given by \autoref{eq:res} as the rotation rate and orbital parameters evolve (see \autoref{fig:e0.5}), although they evaporate for sufficiently small eccentricity. \autoref{eq:res} is even relatively robust to changes in internal structure (save for the complete disappearance of a convective envelope), since inertial waves' frequencies are determined primarily by the Coriolis force \citep[this is also the case for Rossby modes;][]{Papaloizou2023}. 

\subsection{Mode amplitudes during torque balance}
In Section \ref{sec:num} we demonstrate quantitatively that synchronization traps near the period ratios given by \autoref{eq:res} can impede progression toward pseudosynchronous rotation in stars and planets with substantial convective envelopes. First, though, in this section we explore predictions for the amplitudes achieved by modes involved in such a torque balance. Estimates of mode amplitudes are necessary to determine whether or not the linear approximation assumed throughout this paper will remain reasonable throughout tidal evolution. 

\autoref{eq:res} provides only a rough estimate of the period ratios where we expect to find equilibrium rotation states; to estimate the mode amplitudes required to trap the primary's rotation rate, we search for states in which $\dot{\Omega}=0$ due to a balance between the torque contribution associated with one mode (labeled by $\alpha'$) and torques attributed to other processes. \autoref{eq:Odot} and \autoref{eq:kapp} then imply that during such a balance, the frequency detuning associated with the mode $\alpha'$ satisfies
\begin{equation}\label{eq:Dl}
    \frac{\Delta_{\alpha'j}^2}{\gamma_j^2}
    \approx
    -\frac{2\pi}{5}
    \frac{|X_{22j}|^2|Q_{22}^{\alpha'}|^2}{\epsilon_{\alpha'}\tau_f\gamma_j}
    -1,
\end{equation}
where $\tau_f=\sum_k|X_{22k}|^2\kappa_{22k}^f$ is the contribution from nonresonantly driven f-modes (in principle, $\tau_f$ could be modified to include torques from other processes such as magnetic braking), and $j$ is the integer such that $|\Delta_{\alpha'j}|=|\omega_{\alpha'}- \omega_j|\ll1$ is smaller than the detuning associated with any other $k$-harmonic (for the mode $\alpha'$). For this quadratic equation to have real roots, $\epsilon_{\alpha'}\tau_f\gamma_j$ must be both negative and small enough for the first term on the right-hand side to be larger than one. The dimensionless mode amplitude achieved during this balance can be estimated from \citep[e.g.,][]{Dewberry2024}
\begin{equation}
    c_{\alpha'}
    \approx\frac{-\tilde{A}_{22j}Q_{22}^{\alpha'}}{2\epsilon_{\alpha'}\gamma_j
    (\Delta_{\alpha'j}/\gamma_j - \text{i})},
\end{equation}
where $\tilde{A}_{22j}=A_{22j}/(GM/R),$ and $A_{22j}$ is given by \autoref{eq:Acoeff}. Making use of \autoref{eq:Dl}, we then find
\begin{equation}\label{eq:ca2}
    |c_{\alpha'}|^2
    \approx 
    -\frac{3}{4}
    \left(\frac{M_2}{M}\right)^2
    \left(\frac{R}{a}\right)^6
    \frac{\tau_f}{\epsilon_{\alpha'}\gamma_j}
    =\frac{-\tilde{I}\tilde{\Omega}}
    {2\epsilon_{\alpha'}\gamma_j\tilde{t}_{\Omega f}},
\end{equation}
where $\tilde{I}=I/(MR^2)$ and $\tilde{\Omega}=\Omega/\Omega_d$ are the primary's scaled moment of inertia and rotation rate, $\Omega_d=(GM/R^3)^{1/2}$ is its dynamical frequency, $\tilde{t}_{\Omega f}=t_{\Omega f}\Omega_d$, and 
\begin{equation}
    t_{\Omega f}^{-1}
    =\frac{3}{2}\frac{GMM_2}{aI\Omega}\frac{M_2}{M}
    \left(\frac{R}{a}\right)^5\tau_f
\end{equation}
gives the timescale for spin evolution due solely to nonresonant f-mode driving (i.e., equilibrium tidal torques). 

The mode amplitude required for spin equilibrium therefore scales directly with the nonresonant tidal torque encapsulated in $\tau_f$; in order to balance a stronger nonresonant torque, we might expect that a mode would need to be driven further into resonance. However, \autoref{eq:ca2} also involves an inverse dependence on the damping rate $\gamma_j$, which balances the implicit dependence of $\tau_f$ on internal dissipation (through \autoref{eq:kapp}). Equivalently, smaller damping rates $\gamma_j$ due to weaker dissipation are compensated by longer timescales $t_{\Omega f}$. For spin evolution governed primarily by tides, we find that mode amplitudes in a synchronization trap remain small (see \autoref{fig:amp}) and essentially independent of the assumed viscosity (see \autoref{app:visc}).

\section{Quantitative example}\label{sec:num}
\subsection{Interior models}\label{sec:model}
As a proof of concept, we characterize fully self-consistent tidal spin synchronization in polytropic stellar models with pressure and density related by $P=K\rho^{1+1/n}$, for $n=3/2$. Together with a first adiabatic exponent $\Gamma_1=5/3$, this choice of index $n$ produces models that are neutrally stratified (i.e., fully convective). Numerically computed polytropic models are characterized by dimensionless ``eigenvalues'' $\lambda^2=4\pi GR^2\rho_c^{1-1/n}/[K(1+n)],$ where $\rho_c$ is the central density \citep[e.g.,][]{Boyd2011,Dewberry2022b}. To construct a sequence of rotating models representative of solar-mass stars very early on the pre-main sequence, we first choose a primary mass $M=1M_\odot$ and a nonrotating radius of $R=2R_\odot$. This fixes $K$, through $\lambda$. Assuming that this constant remains unchanged with changes in spin, the equatorial radius for a model with a given rotation rate can then be computed from
\begin{equation}
    R=\left[
        \frac{\lambda^2 K(1+n)}{4\pi G}\left(\frac{\hat{M}}{M}\right)^{1-1/n}
    \right]^{n/(3-n)},
\end{equation}
where $\hat{M}=M/(\rho_c R^3)$ is a dimensionless mass. \autoref{app:model} provides further details about models computed with rotation rates up to nearly the dynamical frequency $\Omega_d=(GM/R^3)^{1/2}$, and compares the internal structures recovered to that of a $1M_\odot$, pre-main sequence stellar model computed with MESA \citep{Paxton2011, Paxton2013, Paxton2015, Paxton2018, Paxton2019, Jermyn2023}.

\subsection{Tidal calculations}\label{sec:tide}
We compute the dissipative coefficients $\kappa_{22k}$ for our sequence of rotating stellar models using an aggressively truncated expansions in $m=2$ modes that consist solely of the prograde and retrograde sectoral ($\ell\approx m=2$) f-modes, and the two (prograde and retrograde) $m=2$ inertial modes with the longest wavelengths \citep[see Figure 1 in ][]{Dewberry2022}. \autoref{app:mode} describes the numerically computed mode properties that enter into the calculation, and compares profiles of $\kappa_{222}=\Im[k_{22}(2\Omega_o)]$ computed for our sequence of rotating polytropes (via \autoref{eq:kapp}) to direct tidal calculations for a MESA model of a $1M_\odot$ star on the pre-main sequence.

\subsection{Secular evolution integration}
Adopting the stellar and tidal models described in Section \ref{sec:model} and Section \ref{sec:tide}, we integrate \autoref{eq:Odot} while assuming a fixed orbital eccentricity and semimajor axis. \autoref{eq:Odot} can be written in dimensionless form as
\begin{align}\label{eq:odotnd}
    \frac{\partial\tilde{\Omega}}{\partial \tilde{t}}
    &=\frac{3q^2}{2\tilde{I}\tilde{a}^6}
    \sum_{k=-\infty}^\infty
    |X_{22k}|^2\kappa_{22k},
\end{align}
where again $\tilde{t}=t\Omega_d,$ $\tilde{\Omega}=\Omega/\Omega_d,$ $\tilde{I}=I/(MR^2),$ and $\tilde{a}=a/R.$ Note that $\tilde{I}$ and $\tilde{a}$ are both functions of $\Omega$ (even for a fixed physical semimajor axis), since the equatorial radius $R=R(\Omega)$. We solve \autoref{eq:odotnd} using the Gragg-Bulirsch-Stoer integrator included in REBOUND \citep{rebound,reboundias15}, self-consistently computing $\tilde{I},$ $\tilde{a}$, and $\kappa_{22k}$ for all $k$ such that $|X_{22k}|^2>10^{-6}$ (decreasing this tolerance to $10^{-10}$ had no effect on our results); at each time-step, we interpolate from the sequence of model and mode data described in \autoref{app:model} and \autoref{app:mode} (respectively) in order to generate the coefficients appearing on the right-hand side of \autoref{eq:odotnd}. 

\subsection{Results}
\begin{figure*}
    \centering
    \includegraphics[width=\textwidth]{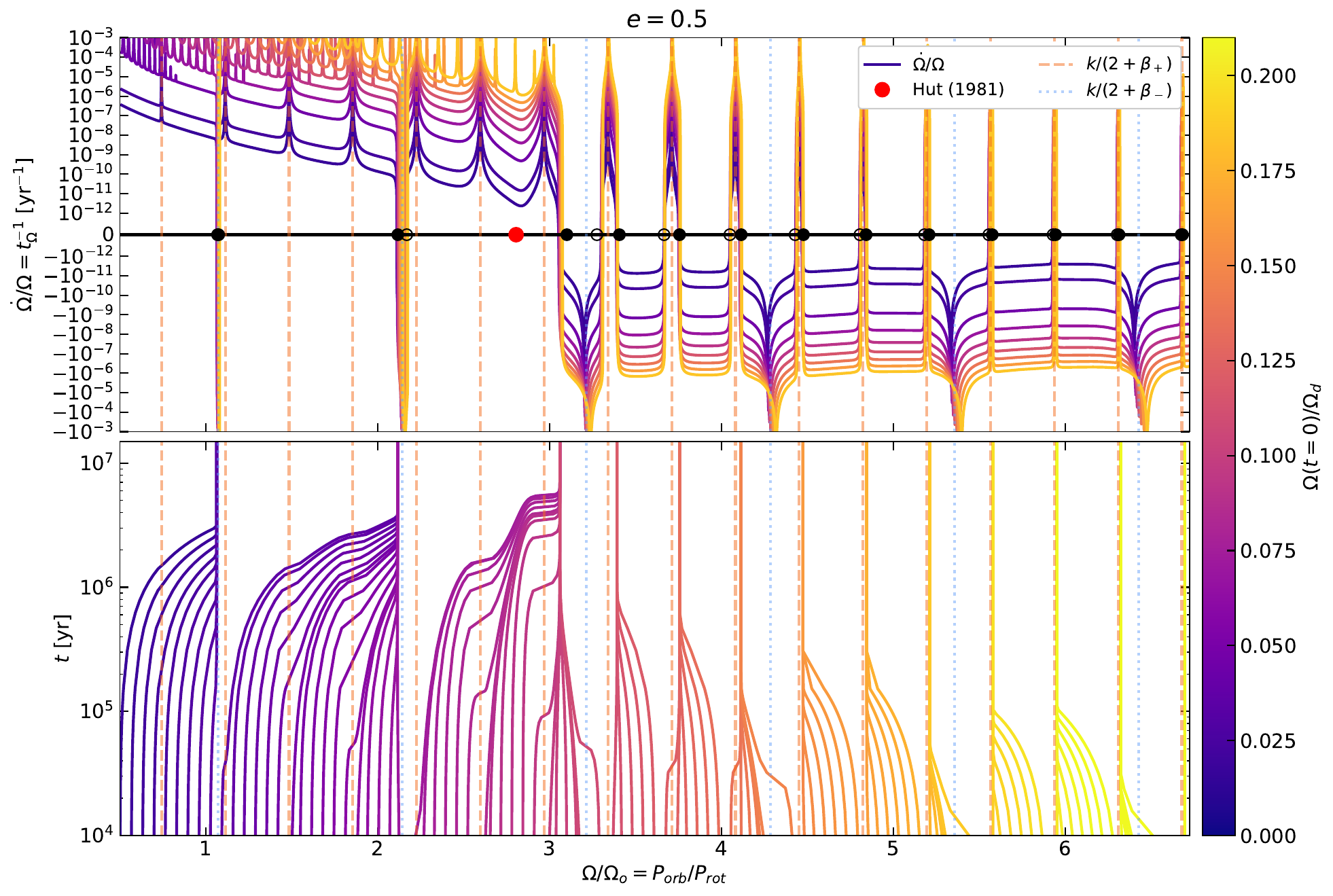}
    \caption{\textbf{Top:} similar to \autoref{fig:demo}, the solid curves plot inverse spin evolution timescales for an $n=1.5$ polytropic, $1M_\odot$ star (with a radius $R=2R_\odot$ in the limit of zero rotation, and an Ekman number of $E_k=10^{-5}$), which is tidally interacting with an equal-mass companion with an eccentricity of $e=0.5$ (the y-axis switches from log to linear scale at $10^{-12}$). The black points now show the zero-crossings in $\dot{\Omega}$ predicted by \autoref{eq:Dl}, and the red point indicates the pseudosynchronous prediction of \citet{Hut1981}. From dark to light, the colormap indicates increasing rotation rate $\Omega$. Because inertial-wave frequencies scale directly with $\Omega$, the values of $\Omega/\Omega_o$ where $\dot{\Omega}$ vanishes do not vary with changes in rotation rate. \textbf{Bottom:} ratios $\Omega/\Omega_o$ (x-axis) as a function of time (y-axis) computed by directly integrating \autoref{eq:odotnd} for a fixed $P_\text{orb}=10$ days, and initial rotation rates $\Omega$ indicated by the color-scale. The period ratio-invariant zero-crossings in $\dot{\Omega}$ due to inertial modes (top panel) impede the star's progression toward the pseudosynchronization.}
    \label{fig:e0.5}
\end{figure*}

\autoref{fig:e0.5} illustrates the spin evolution of a star in an equal-mass system with (i) an eccentricity of $e=0.5$, (ii) a constant Ekman number of $E_k=\nu/(R^2\Omega)=10^{-5}$ (here $\nu$ is a kinematic viscosity), and (iii) rotation rates (indicated by the colorbar) ranging from $\Omega\approx0.02\Omega_d$ to $0.23\Omega_d$ (rotation periods ranging from $P_\text{rot}\approx20.5$ days to $1.5$ days). The curves in the top panel plot the inverse of rotational evolutionary timescales $t_\Omega$ defined by $t_\Omega^{-1}=\dot{\Omega}/\Omega$ (in units of inverse years) as a function of $\Omega/\Omega_o=P_\text{orb}/P_\text{rot}$, with darker (lighter) colors indicating slower (faster) rotation rates $\Omega$. Zero-crossings of these curves delineate equilibrium states with $\dot{\Omega}=0$. The red dot in \autoref{fig:e0.5} (top) indicates the pseudosynchronous equilibrium state predicted by \citet{Hut1981}.

As discussed in Section \ref{sec:eqm}, calculations including only f-modes (see the dark blue line in \autoref{fig:demo}) predict vanishing torques at only a single period ratio that generally falls close to this prediction, except when rotation becomes fast enough and/or eccentricity large enough for f-modes to be driven resonantly. The inclusion of the two longest-wavelength $m=2$ inertial modes leads to more interesting predictions for spin evolution, producing numerous zero-crossings in $\dot{\Omega}$ close to the ratios of \autoref{eq:res} (indicated by orange dashed lines for the prograde inertial mode, and blue dotted lines for the retrograde mode). The black points indicate predictions for the zero-crossings close to these ratios, computed using the frequency detuning estimate given by \autoref{eq:Dl}. Small deviations between these predictions and the actual zero-crossings of $\dot{\Omega}$ occur when the f-mode torque is balanced by more than the driving of an inertial mode by a single Fourier component of the tidal potential. 

Importantly, the scaling of inertial mode frequencies with rotation rate ensures that these zero-crossings occur at ratios of orbital to rotation period that do not shift with changing rotation period (i.e., the light and dark colored lines cross zero at the same values of $\Omega/\Omega_o$, regardless of the individual values of $\Omega$ or $\Omega_o$). The consequence of this commonality is that every ratio $\Omega/\Omega_o=k/(2+\beta_\pm)$ involving an inertial mode driven strongly enough by the $k$'th tidal component to balance the nonresonant (f-mode) torque introduces a pair of zero-crossings in $\dot{\Omega}$ that constitute fixed points in the spin evolution. 

The fixed points involving a negative slope in $\dot{\Omega}/\Omega$ with respect to $\Omega/\Omega_o$ are stable (since a positive $\dot{\Omega}$ will push the spin evolution toward larger values of $\Omega$; see the arrows in \autoref{fig:demo}), and the bottom panel of \autoref{fig:e0.5} illustrates the efficacy of these stable fixed points in ``trapping'' the primary body's rotation rate. The curves in this panel plot the same ratio $\Omega/\Omega_o$ (x-axis), computed as a function of time (y-axis) by integrating \autoref{eq:odotnd} while assuming a fixed orbital period of $10$ days. These curves indicate first of all that the torques plotted in the top panel act to drive the star's spin toward the pseudosynchronous state predicted by \citet{Hut1981}. Along the way, though, the inertial mode resonances invariably halt progression toward pseudosynchronization at ratios of $\Omega/\Omega_o$ just above (below) $k/(2+\beta_\pm)$ if the nonresonant f-mode torque is negative (positive). 

The time taken for rotation rates to become stuck in one of these synchronization traps decreases with increasing eccentricity (see \autoref{app:evar}), and with increasing viscosity (see \autoref{app:visc}). Because we have for simplicity assumed a constant Ekman number $E_k=\nu/(R^2\Omega),$ the latter point explains why \autoref{fig:e0.5} indicates more rapid spin evolution in stars that are initially spinning more rapidly. The ability of an inertial mode resonance to act as a synchronization trap is relatively insensitive to the nature and amplitude of viscous dissipation, though, requiring only that the near-resonant response be large enough to balance the (typically small) nonresonant f-mode torque. Indeed, we find that the period ratio where $\dot{\Omega}$ vanishes (as determined from \autoref{eq:Dl}, or from the numerically computed curves) is essentially invariant with Ekman number. This invariance (demonstrated by \autoref{fig:wvar}) derives from the fact that although tidal resonances generally become narrower with decreasing viscosity (see \autoref{fig:k22vs}), a more weakly damped inertial mode only has to balance the nonresonant f-mode torque (which also weakens with decreasing viscosity).

\begin{figure*}
    \centering
    \includegraphics[width=\textwidth]{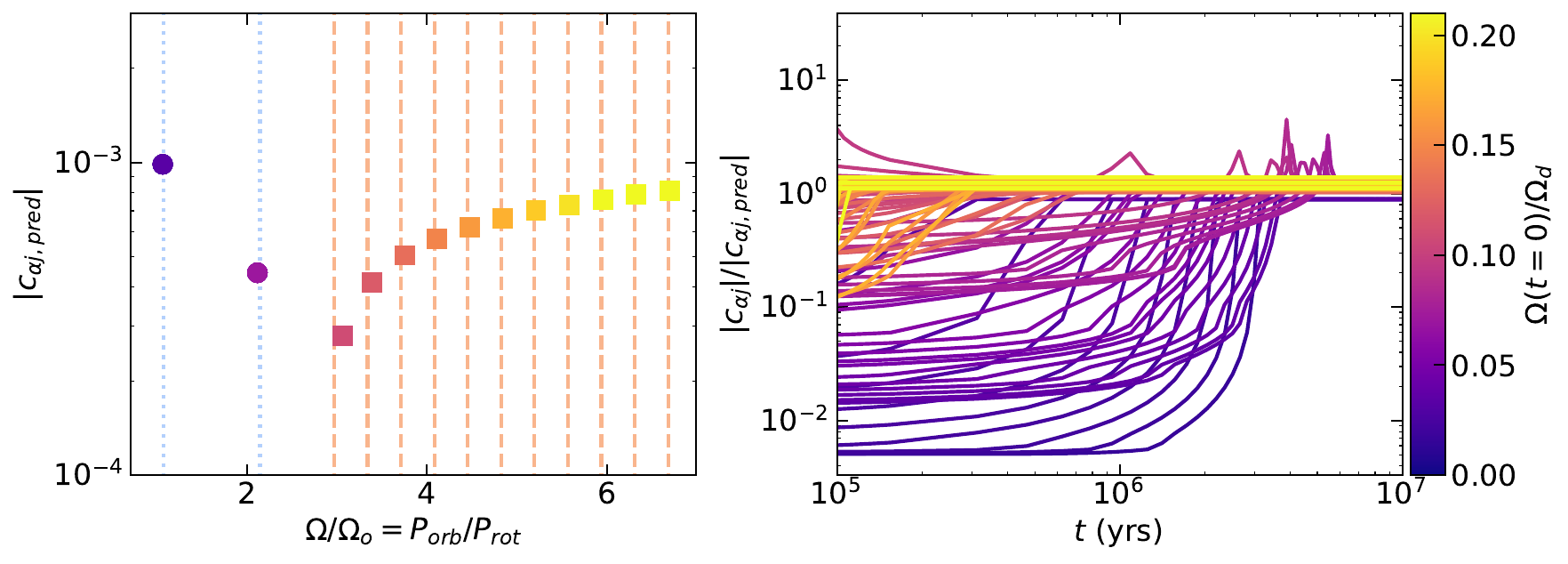}
    \caption{\textbf{Left:} dimensionless amplitudes predicted by \autoref{eq:ca2} for the end states of all of the integrations shown in \autoref{fig:e0.5} (bottom). Circles (squares) indicate spins trapped by torque balances due to the retrograde (prograde) inertial mode (because the integrations all end up with trapped spins, many points lie on top of one another). \textbf{Right:} mode amplitudes computed as a function of time for the numerical integrations shown in \autoref{fig:e0.5} (bottom), normalized by the end-state predictions shown in the left panel. The curves demonstrate that (i) \autoref{eq:ca2} accurately predicts the amplitudes of modes involved in a synchronization trap, and (ii) these amplitudes remain linear.}
    \label{fig:amp}
\end{figure*}

\begin{figure}
    \centering
    \includegraphics[width=\columnwidth]{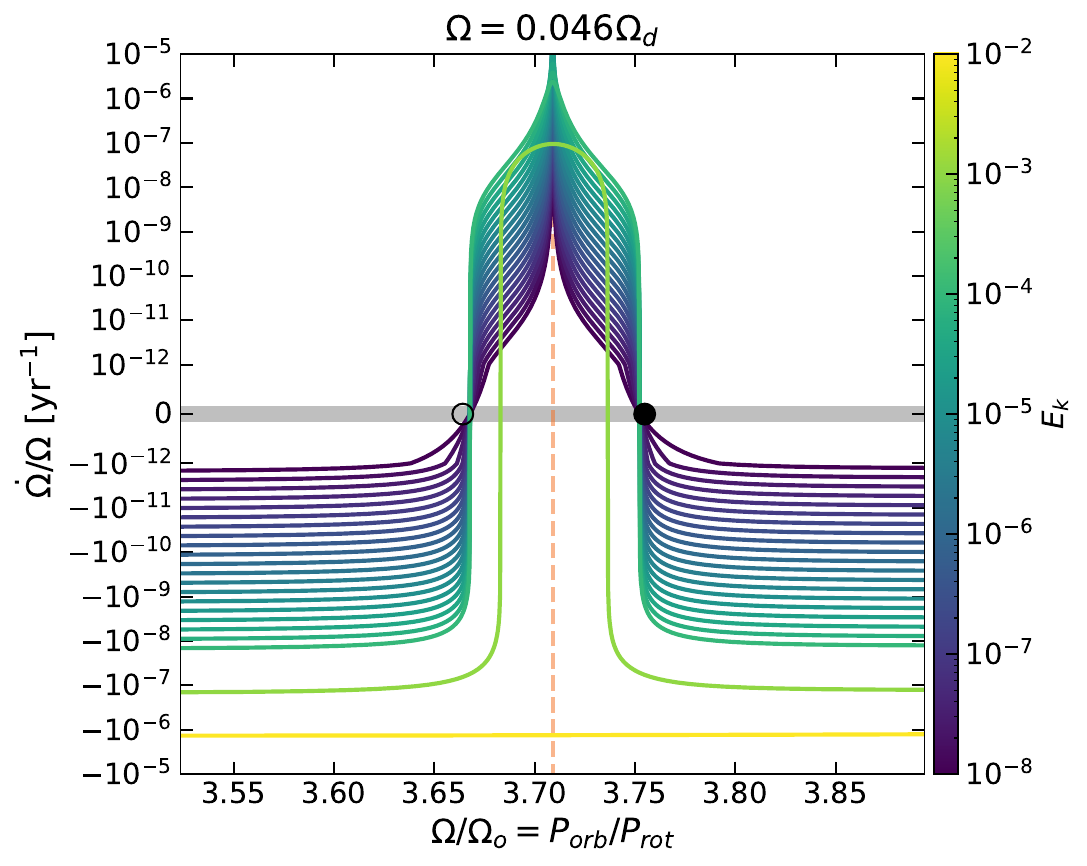}
    \caption{Inverse spin synchronization timescales computed for a model with fixed $\Omega\approx0.046\Omega_d$ and varying Ekman number (indicated by color-scale), plotted as a function of varying period ratios close to the $k=10$ synchronization trap involving the prograde inertial mode. The black points again plot the predictions of \autoref{eq:Dl} in the lowest-viscosity case. These curves demonstrate that the period ratios where the total torque vanishes do not vary with viscosity, except when that viscosity is very large.}
    \label{fig:wvar}
\end{figure}

Because the torque balance required for a synchronization trap does not drive inertial modes deeply into resonance, mode amplitudes need not be large. \autoref{fig:amp} (left) plots the amplitudes predicted by \autoref{eq:ca2}, computed from values of $\tau_f$ and $R$ evaluated at the end of each of the integrations shown in \autoref{fig:e0.5} (bottom). The right-hand panel then compares these predictions to the mode amplitudes actually computed during the integrations, demonstrating agreement within $\sim10\%$. For the long-wavelength inertial modes considered here, the dimensionless amplitudes of $|c_{\alpha}|\sim10^{-4}$ to $10^{-3}$ shown in \autoref{fig:amp} correspond to radial derivatives of the radial Lagrangian displacement of $|\partial \xi_r/\partial r|\lesssim10^{-3}$ 
\citep[$|\partial \xi_r/\partial r|\gtrsim1$ is typically taken as a criterion for wave breaking, since it indicates fluid velocities in excess of the wave's phase speed;][]{Wienkers2018}. 

In this respect synchronization trapping differs from resonance locking \citep[e.g.,][]{Fuller2017}, which involves continuous resonant tidal wave excitation that can often drive modes to nonlinear amplitudes \citep{Ma2021}; our spin equilibria are determined by a balance that causes the total torque to vanish, rather than a continuous transfer of energy into mode driving. Spin regulation by synchronization trapping also promises to be more or less automatic (barring extreme changes in internal structure), since it does not rely on the coincident evolution of tidal and mode frequencies (just the continued existence of a large convective region). 

\section{Discussion}\label{sec:disc}
\subsection{Astrophysical relevance}
The astrophysical relevance of the physical mechanism described in this paper depends on the timescales over which tidal synchronization takes place, which in turn depend on the highly uncertain interaction of tidal and convective flows. By adopting a constant Ekman number in our calculations, we omit a reduction in the efficiency of convective damping that is expected with increasing tidal frequency \citep{Zahn1989,Goodman1997,Duguid2020}, as well as the question of whether convective turbulence should even be expected to act as an effective viscosity when tidal frequencies outstrip convective turnover times \citep{Terquem2021,Barker2021,Terquem2023}. Similarly, the effect of tides on spin rate may be overcome by additional processes \citep[like magnetic braking; ][]{Matt2015} in many systems, although additional weak torques may only modify the balance between the mode resonance and the nonresonant f-mode torque (i.e., change the location of zero-crossings in $\dot{\Omega}$). Additionally, our assumption that the tidally perturbed body spins as a rigid body is oversimplified \citep{Goldreich1989,Astoul2022}.

However, we contend that---in the event that tidal dissipation does act in eccentric systems to regulate the spins of stars or planets with considerable convective envelopes---synchronization traps produced by inertial modes will hinder progression toward pseudosynchronization. Rossby modes \citep{Papaloizou2023} might also play a similar role in bodies with significant stable stratification, although their ability to regulate spin evolution will likely be outweighed by resonance locks involving gravito-inertial oscillations with frequencies that do not scale directly with the rotation rate, and hence do not attract rotation rates to period ratios that are as simple to predict \citep{Witte1999,Witte2001}.

We consider fully convective models as an approximation to solar-mass stars on the pre-main sequence. The dynamical tidal response will of course vary as the radiative core in such stars grows, but \citet{Lin2021} demonstrated that the dominant dissipation in models with convective envelopes and rigid cores is still associated with long-wavelength flows closely resembling the inertial modes of fully isentropic bodies (in both spatial structure and frequency). We therefore expect the synchronization traps associated with the two longest-wavelength inertial modes to remain relevant. However, both radiative and solid cores generically produce enhanced inertial wave dissipation at many more frequencies within the range $[-2\Omega,2\Omega]$ than are relevant in fully convective models \citep{Ogilvie2009,Ogilvie2013,Lin2023,Dewberry2023,Pontin2023}. The presence of a core may therefore lead to synchronization trapping within an even more granular set of spin equilibria. 

How spins evolve within the network of rotation equilibria described in this paper will in turn affect the dominant channels for tidal damping of semimajor axes, eccentricities, and spin-orbit misalignments, by altering the tidal frequencies felt in the corotating frame of the fluid \citep{Ogilvie2007}. Frequency-dependent spin regulation by inertial and/or Rossby waves may therefore be important to consider in attempting to explain the circularization of solar-type binaries \citep{Meibom2005,Zanazzi2022}. 

\subsection{Comparison with observations}

\begin{figure}
    \centering
    \includegraphics[width=\columnwidth]{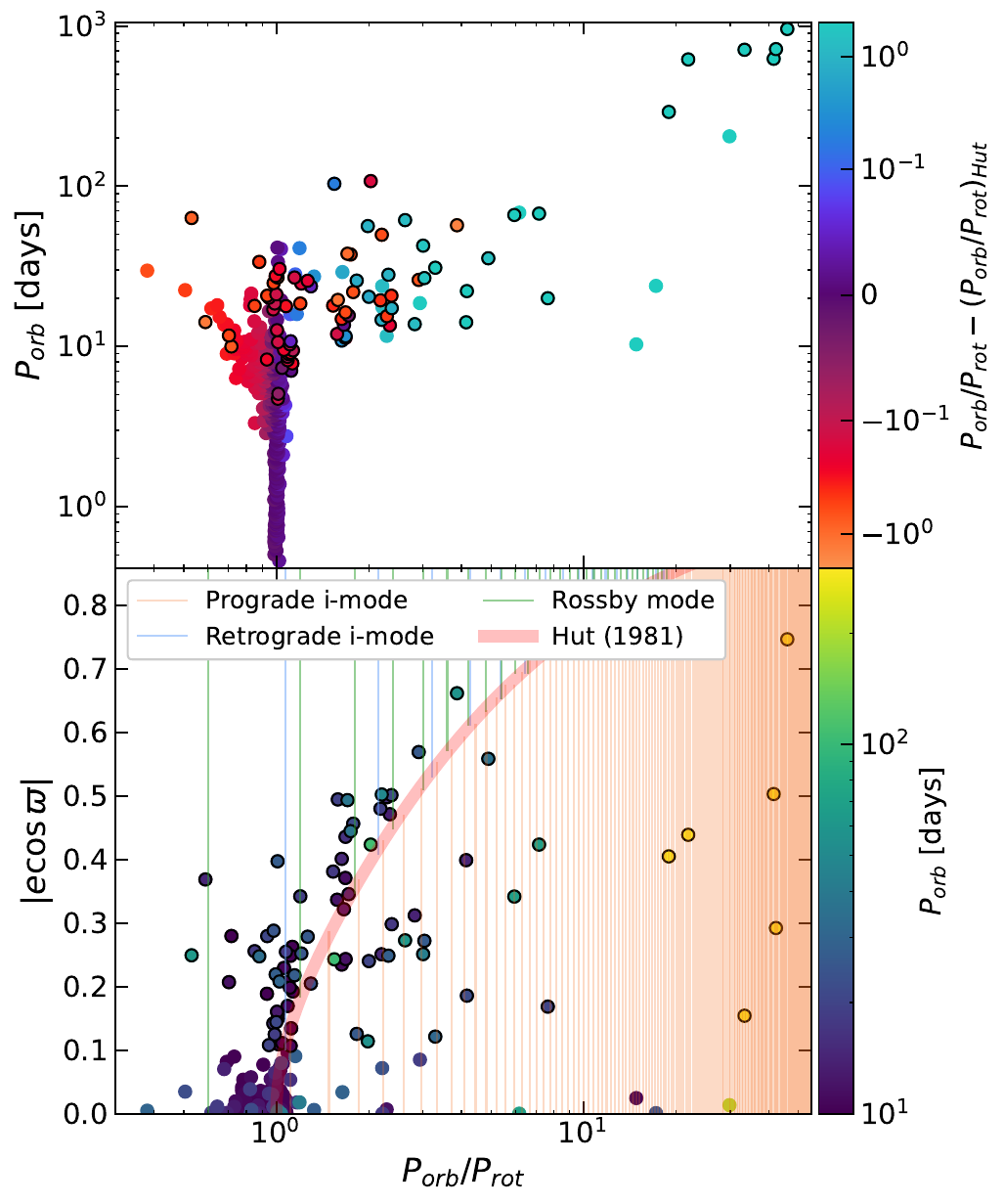}
    \caption{Orbital periods (top) and eccentricity estimates (bottom), plotted as a function of period ratio for the sample of eclipsing binaries considered by \citet{Lurie2017}. In the top panel, point colors indicate departures from pseudosynchronization (assuming  eccentricities $e\sim |e\cos\varpi|$ from the bottom panel). Point colors in the bottom panel indicate orbital period, and in both panels black outlines indicate systems with $|e\cos\varpi|\gtrsim0.1$. Discrepancies from the pseudosynchronous prediction of \citet{Hut1981} could be explained by the synchronization traps indicated by the vertical lines in the bottom panel.}
    \label{fig:ebs}
\end{figure}

Observations offer some evidence that pseudosynchronization may not operate unimpeded in nature. \autoref{fig:ebs} plots the period ratios of eclipsing binaries obtained by \citet{Lurie2017} against orbital periods (top panel) and eccentricity estimates (bottom panel). We computed these eccentricities by cross-referencing with the Villanova Kepler Eclipsing Binary Catalog \citep{Conroy2014}, and using the relation $e\cos\varpi=(\pi/2)(s - 0.5),$ where $\varpi$ is the pericentre angle relative to line-of-sight, and $s=(t_s-t_p)/P_\text{orb}$ is the difference between primary and secondary eclipses \citep{Zanazzi2022}. Taking these estimates of $|e\cos\varpi|$ as an approximation of (in fact a lower bound for) $e$, point colors in the top panel indicate deviations from the pseudosynchronization expected for a given eccentricity. Although the data cluster around the pseudosynchronous prediction given by \autoref{eq:hut} (red line in the bottom panel), they also exhibit significant departures from this prediction. These deviations could simply originate in binaries that have not had enough time to synchronize. However, the physical mechanism explored in this paper also predicts that trajectories toward pseudosynchronization in eccentric systems would be impeded by the numerous synchronization traps indicated by the vertical lines in the bottom panel. 

\begin{figure*}
    \centering
    \includegraphics[width=\textwidth]{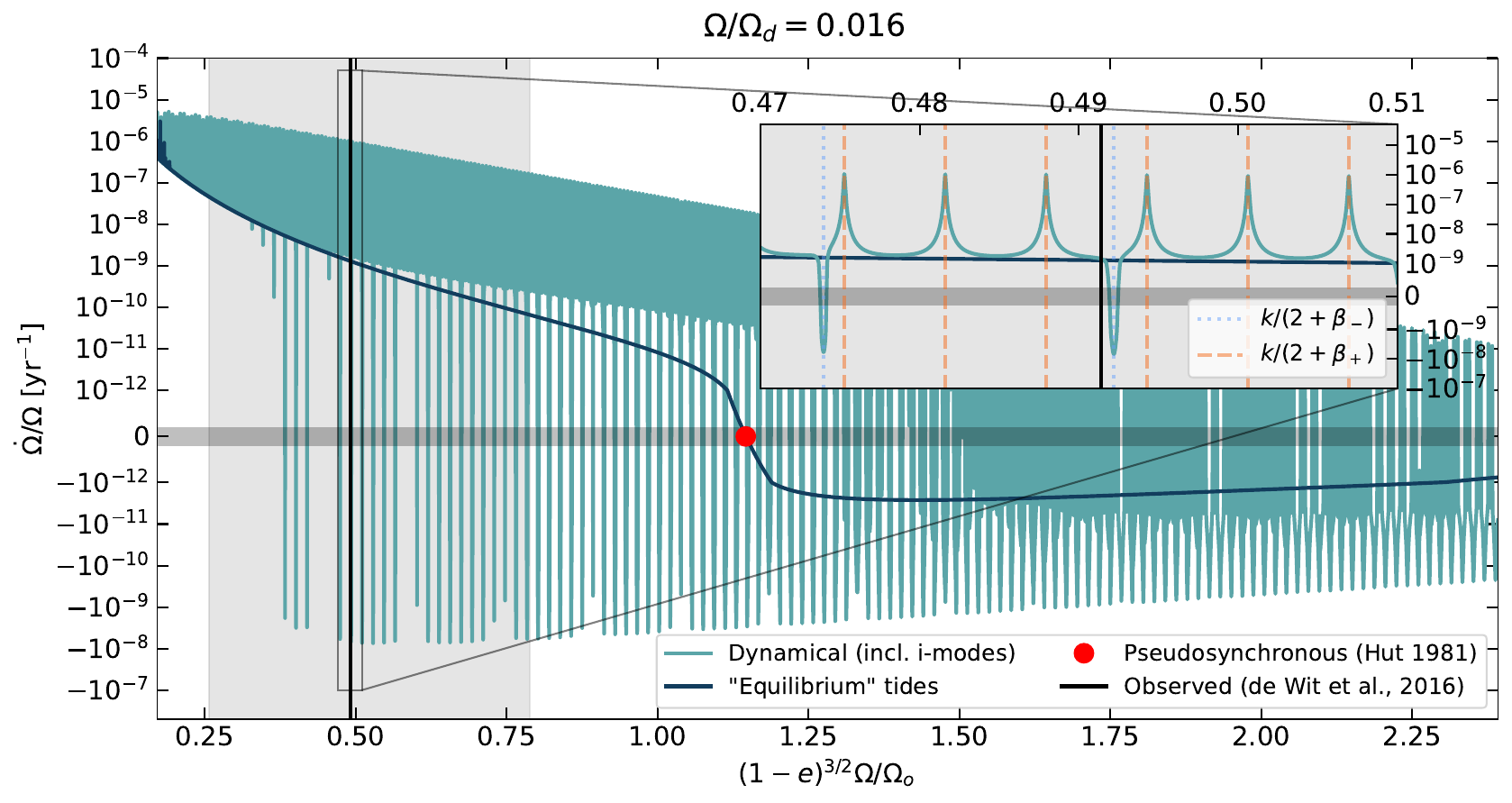}
    \caption{Inverse spin evolution timescales (right-hand side of \autoref{eq:Odot}) computed for parameters roughly appropriate to the hot Jupiter HD 80606 b (we assume a rotation rate $\Omega/\Omega_d\approx0.016,$ mass ratio $M_*/M_p=M_\odot/(3.94M_J)\approx266,$ a planetary radius $R_p=R_J$, an eccentricity $e=0.93366$, and an Ekman number $E_k=10^{-5}$). As in \autoref{fig:demo}, the light and dark blue curves plot profiles of $\dot{\Omega}/\Omega$ computed both with and without the dynamical tidal effects of inertial modes (respectively). Due to the high eccentricity, we plot these curves as a function of $(1-e)^{3/2}\Omega/\Omega_o$ rather than the period ratio. The red point indicates the pseudosynchronous prediction of \citet{Hut1981}, while the vertical black line and gray shading indicate the period ratio and uncertainties reported by \citet{deWit2016}. As shown in the inset, this asynchronous observation could be explained by numerous synchronization traps produced where the retrograde inertial mode causes $\dot{\Omega}/\Omega$ to cross zero.}
    \label{fig:HD}
\end{figure*}

In stars hosting exoplanets, stellar rotation rates are broadly distributed over a range of asynchronous period ratios \citep{Bruno2023}. Synchronization trapping likely is not needed to explain this asynchronicity, though, since many exoplanets are not massive enough to efficiently regulate the spins of their host stars. The exoplanets themselves, on the other hand, ought to be spun up or down more efficiently. The hot Jupiter HD 80606 b \citep{deWit2016} is highly eccentric, and possesses a spin period that deviates significantly from the expected pseudosynchronous rotation rate. Assuming that HD 80606 b is largely convective and possesses inertial modes with frequencies that scale similarly to the models considered here, the observed period ratio of $P_\text{orb}/P_\text{rot}\sim29$ could be explained by a synchronization trap involving the retrograde inertial mode and $k\sim25-30$. \autoref{fig:HD} plots spin evolution timescales computed for a polytropic model with parameters scaled to match those of HD 80606 b. The spin evolution profiles shown in \autoref{fig:HD} involve a dense forest of resonances, owing to the system's high eccentricity. The retrograde inertial mode generically produces synchronization traps that coincide with the observed period ratio constraints.

\section{Summary}\label{sec:sum}
We have demonstrated that in stars and planets with large convective regions, tidal driving of inertial oscillations by an eccentric orbit can introduce numerous spin equilibria beyond the pseudosynchronous state associated with constant time-lag or equilibrium tidal models \citep{Hut1981}. A direct consequence of the fact that inertial-wave frequencies scale with the rotation rate is that a subset of these additional equilibria act as stable fixed points at particular values (given by \autoref{eq:res}) of the period ratio $P_\text{orb}/P_\text{rot}$. They can therefore inhibit progression toward pseudosynchronization (a process commonly assumed to take place rapidly in tidally interacting systems), ``trapping'' spins at values that are closer to initial rotation rates. 

The mechanism underlying this process of synchronization trapping is similar to resonance locking, except that near-resonances are maintained automatically by the scaling of inertial waves' frequencies with the rotation rate. Additionally, we find that the balance of torques does not drive inertial mode amplitudes to nonlinearity, and that the mode amplitudes and frequency detunings are insensitive to the effective viscosity ascribed to convective turbulence (at least in situations where tides dominate spin evolution). 

The calculations presented in this paper are simplified, but find motivation in observations of both stellar binaries and exoplanet systems with rotation rates that do not conform to pseudosynchronous expectations. We also expect synchronization trapping to affect subsequent damping of orbital eccentricities, semimajor axes, and spin-orbit misalignments, by altering the frequencies of tidal driving. Future studies should investigate the interaction of the spin regulation considered here with the secular evolution of the other orbital elements. 

%% Also note that the akcnowlodgment environment does not support long amounts of text. If you have a lot of people and institutions to acknowledge, do not use this command. Instead, create a new \section{Acknowledgments}.

\ack{I thank the reviewer for constructive comments that improved the quality of this paper. I also thank Yanqin Wu, Jim Fuller, and J. J. Zanazzi for helpful conversations. This work was supported by the Natural Sciences and Engineering Research Council of Canada (NSERC) [funding reference \#CITA 490888-16].}
% \begin{acknowledgments}
% \end{acknowledgments}


% \vspace{5mm}
% \facilities{}


\software{numpy \citep{numpy}, matplotlib \cite{matplotlib}, REBOUND \citep{rebound}, 
MESA \citep{Paxton2011, Paxton2013, Paxton2015, Paxton2018, Paxton2019, Jermyn2023}, CMasher \citep{cmasher}
}



\appendix

\section{Tidal evolution equations}\label{app:teveqn}
\subsection{Tidal and response potentials}\label{app:pot}
Working with spherical coordinates $(r,\theta,\phi)$ in an inertial frame with the origin fixed at the center of a tidally disturbed fluid body that is axisymmetric in isolation (with mass $M$ and equatorial radius $R$), consider the gravitational potential $U=U({\bf r},t)$ of a tidal perturber with separation ${\bf d}(t)$, mass $M_2$, and negligible spin-orbit misalignment. Assuming that this perturber is sufficiently distant that it can be treated as a point-mass, the tidal potential can be expanded in spherical harmonics as    
\begin{equation}\label{eq:Uexp}
    U=\frac{-GM_2}{|{\bf r} - {\bf d}|}
    =\Re \left\{
        \sum_{n=2}^\infty
        \sum_{m=0}^n
        \sum_{k=-\infty}^\infty
        A_{nmk}\left(\frac{r}{R}\right)^n
        Y_{nm}(\theta,\phi)
        \exp[-\text{i}k\Omega_ot]
    \right\}.
\end{equation}
Here $Y_{nm}$ are spherical harmonics, $\Omega_o=[G(M+M_2)/a^3]^{1/2}$ is the mean motion, $a$ is the orbital semimajor axis, and we omit the $n=0,1$ terms responsible for basic Keplerian orbital motion \citep[e.g.,][]{Ogilvie2014}. The coefficients $A_{nmk}$ are given by 
\begin{equation}\label{eq:Acoeff}
    A_{nmk}=-c_{mk}\left(\frac{GM_2}{a}\right)
        \left(\frac{R}{a}\right)^n
        \left(\frac{4\pi }{2n+1}\right)
        Y_{nm}^*(\pi/2,0)X_{nmk}(e),
\end{equation}
where 
\begin{equation}
    c_{mk}
    =\begin{cases}
        0, & m=0,k<0,\\
        1, & m=0, k=0,\\
        2, & \text{otherwise}.
    \end{cases}
\end{equation}
Here $e$ is the orbital eccentricity, and
\begin{equation}\label{eq:hansen}
    X_{nmk}
    =\frac{1}{\pi}
    \int_0^{\pi}(1 - e\cos E)^{-n}
    \cos\Bigg[ 
        2m\tan^{-1}\left(\sqrt{\frac{1+e}{1-e}}\tan \frac{E}{2}\right)
        -k(E-e\sin E)
    \Bigg]
    \text{d}E
\end{equation}
are Hansen coefficients. We truncate the tidal potential at the quadrupolar spherical harmonic degree $n=2$.

The tidal force $-\nabla U$ induces fluid motions, density variations, and consequently an Eulerian perturbation $\delta\Phi({\bf r},t)$ to the equilibrium potential $\Phi_0({\bf r})$ of the tidally perturbed body. The gravitational perturbation can similarly be expanded as
\begin{equation}\label{eq:dPexp}
    \delta\Phi 
    =\Re \left\{
        \sum_{\ell=0}^\infty\sum_{m=0}^\ell\sum_{k=-\infty}^\infty
        \Phi'_{\ell m k}\left(\frac{R}{r}\right)^{\ell+1} 
        Y_{\ell m}\exp[-\text{i}k\Omega_ot]
    \right\}.
\end{equation}
We assume that the coefficients $\Phi'_{\ell mk}$ and $A_{nmk}$ of a given $k$ and $m$ can be linearly related via (complex-valued) potential Love numbers $k_{\ell m}^n=k_{\ell m}^n(k\Omega_o)$:
\begin{equation}
    \Phi'_{\ell m k}
    =\sum_{n=2}^\infty k_{\ell m}^nA_{nmk}.
\end{equation}
For our purely quadrupolar tidal potential, this linear relation reduces to $\Phi'_{\ell mk}=k_{\ell m}^2A_{2mk}.$

\subsection{Tidal power and torque}
The rate of energy transfer in our inertial frame is given by \citep{Ogilvie2013}
\begin{equation}
    \frac{1}{4\pi G}
    \int (
        \delta\dot{\Phi}\nabla U
        -U\nabla\delta\dot{\Phi}
    )\cdot\text{d}{\bf S},
\end{equation}
where the surface integral is carried out on any spherical surface of radius $s>R$ that contains the primary and excludes the secondary. Inserting the expansions in \autoref{eq:Uexp} and \autoref{eq:dPexp}, the time-steady rate of energy transfer (tidal power) in the inertial frame can be found as
\begin{equation}
    P=\sum_{\ell=2}^\infty 
    \sum_{m=0}^\ell
    \sum_{k=-\infty}^\infty
    \frac{(2\ell+1)}{8\pi G}R
    k\Omega_o\sum_{n=2}^\infty \Im[k_{\ell m}^n(k\Omega_o)]
    A_{nmk}A_{\ell mk}^*.
\end{equation}
With our drastic truncation of the tidal potential at $n=2$,  the tidal power simplifies to 
\begin{equation}
    P
    =\frac{GM_2^2}{2a}
    \left(\frac{R}{a}\right)^5
    \sum_{k=-\infty}^\infty
    k\Omega_o
    \left(
        \frac{1}{2}
        |X_{20k}|^2\kappa_{20k} 
        +\frac{3}{2}|X_{22k}|^2\kappa_{22k}
    \right),
\end{equation}
where $\kappa_{\ell mk}=\Im[k_{\ell m}(k\Omega_o)]$ gives the imaginary part of the Love number $k_{\ell m}\equiv k_{\ell m}^\ell$ describing  the response in the same harmonic as the driving. 

Similarly, for aligned spin and orbital angular momenta the tidal torque imposed on the perturbed body is given by 
\begin{equation}
    {\bf T}=-\int \delta\rho
    {\bf r}\times\nabla U\text{d}V
    =\left\{\frac{1}{4\pi G}\int [
        \delta\Phi \nabla (\partial_\phi U)
        - (\partial_\phi U)\nabla \delta \Phi 
    ]\cdot\text{d}{\bf S}\right\}\hat{\bf z},
\end{equation}
where $\delta\rho$ is the Eulerian density perturbation. The steady rate of angular momentum transfer due to a quadrupolar tidal potential is then
\begin{equation}
    T
    =\frac{10R}{8\pi G}
    \sum_{k=-\infty}^\infty
    |A_{22k}|^2\kappa_{22k}
    =\frac{3}{2}\frac{GM_2^2}{a}
    \left(\frac{R}{a}\right)^5
    \sum_{k=-\infty}^\infty
    |X_{22k}|^2\kappa_{22k}.
\end{equation}

\subsection{Secular equations}\label{app:sec}
The orbital energy is $E_o=-GMM_2/(2a)$. Writing $\dot{E}_o=-P$ then produces
\begin{equation}\label{eq:adot}
    \frac{\dot{a}}{a} 
    =-\frac{M_2}{M}
    \left(\frac{R}{a}\right)^5
    \sum_{k=-\infty}^\infty
    k\Omega_o
    \left(
        \frac{1}{2}|X_{20k}|^2\kappa_{20k}
        +\frac{3}{2}|X_{22k}|^2\kappa_{22k}
    \right).
\end{equation}

Meanwhile, assuming that the tidally perturbed body rotates rigidly, and equating the rate of angular momentum transfer with the time derivative of the tidally perturbed body's spin angular momentum $J=I\Omega$ (here $I=\int_V \rho_0R^2\text{d}V$ is the moment of inertia) gives
\begin{equation}
    \frac{\dot{\Omega}}{\Omega}
    =\frac{3}{2}\frac{GMM_2}{aI\Omega}\frac{M_2}{M}
    \left(\frac{R}{a}\right)^5
    \sum_{k=-\infty}^\infty
    |X_{22k}|^2\kappa_{22k}.
\end{equation}
Here we have ignored $\dot{I}$, which is technically nonzero for nonzero $\dot{\Omega}$ (since changes in $\Omega$ lead to changes in the equilibrium density distribution $\rho_0$). Finally, writing $T=-\dot{L}_o$, where $L_o=-2(1-e^2)^{1/2}E_o/\Omega_o$ is the orbital angular momentum, produces
\begin{align}\label{eq:edot}
    \frac{\dot{e}}{e}
    &=\frac{(1-e^2)}{e^2}\left( 
        \frac{1}{2}\frac{\dot{a}}{a}
        +\frac{1}{L_o}T
    \right)
\\\notag
    &=\frac{(1-e^2)}{e^2}
    \frac{M_2}{M}
    \left(\frac{R}{a}\right)^5
    \Omega_o\sum_{k=-\infty}^\infty
    \left[
        \frac{3}{4}
        \left(
            %\frac{-2E_o}{-2E_o}
            \frac{2}{(1-e^2)^{1/2}}
            -k
        \right)
        |X_{22k}|^2\kappa_{22k}
        -\frac{1}{4}
        k|X_{20k}|^2\kappa_{20k}
    \right].
\end{align}

\begin{figure*}
    \centering
    \includegraphics[width=\textwidth]{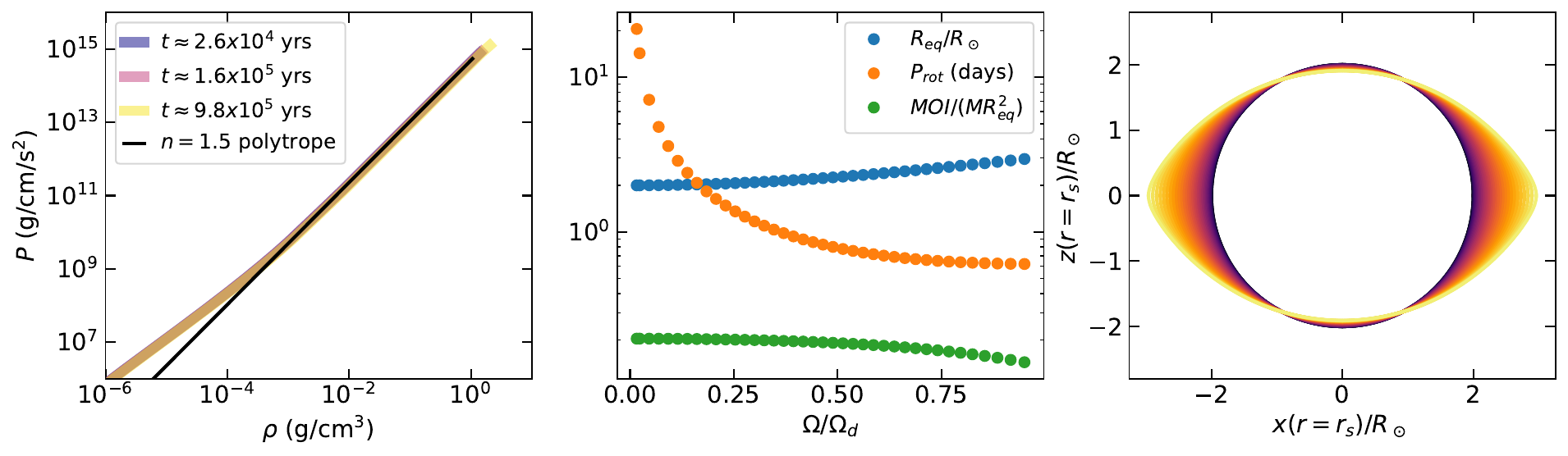}
    \caption{\textbf{Left:} Pressure vs. density, plotted for a nonrotating $n=1.5$ polytropic star with $M=1M_\odot$ and $R=2R_\odot$ (black line), and snapshots of a $1M_\odot$ MESA model on the pre-main sequence (colored lines). \textbf{Middle:} Equatorial radius (blue), rotation period (orange), and dimensionless moment of inertia (green), plotted as a function of $\Omega/\Omega_d$ for the $1M_\odot$, $n=1.5$ polytropic sequence of rotating models incorporated into the integration of \autoref{eq:odotnd}. \textbf{Right:} meridiodonal cuts showing the surface radii of the polytropic sequence, from $\Omega/\Omega_d\approx0.016$ (dark blue) to $\Omega/\Omega_d\approx0.95$ (bright yellow).} 
    \label{fig:model}
\end{figure*}

\section{Rotating model sequence}\label{app:model}
\autoref{fig:model} summarizes the properties of the sequence of rotating models considered in this study. The lefthand panel plots pressure vs. density (in cgs units) for a nonrotating $n=1.5$ polytrope with mass $M=1M_\odot$ and radius $R=2R_\odot$ (black line), and for snapshots of a MESA model of a $1M_\odot$ star during the pre-main sequence (colored lines). The middle panel in \autoref{fig:model} plots equatorial radii (blue), rotation periods (orange), and dimensionless moments of inertia as a function of rotation rate for the sequence of $n=1.5$ polytropes described in Section \ref{sec:model}. Lastly, the meridional cuts in the right-hand panel show the surface radius for each rotating model in the sequence. The sequence of polytropic models demonstrated in \autoref{fig:model} was computed using a Newton-Kantorovich iteration \citep[as described in][]{Boyd2011,Dewberry2022b}.\footnote{
A simple code for computing rotating polytropic models is freely available at \href{https://github.com/dewberryjanosz/polyrot}{https://github.com/dewberryjanosz/polyrot}.
} This approach yields results in agreement with classic calculations of rapidly rotating polytropic models \citep[e.g.,][]{James1964,Ostriker1968,Hachisu1986}. The MESA model was computed with version r15140, using the 
\texttt{1M\_pre\_ms\_to\_wd} 
test suite example. 

\section{Oscillation mode properties}\label{app:mode}
\begin{figure*}
    \centering
    \includegraphics[width=\textwidth]{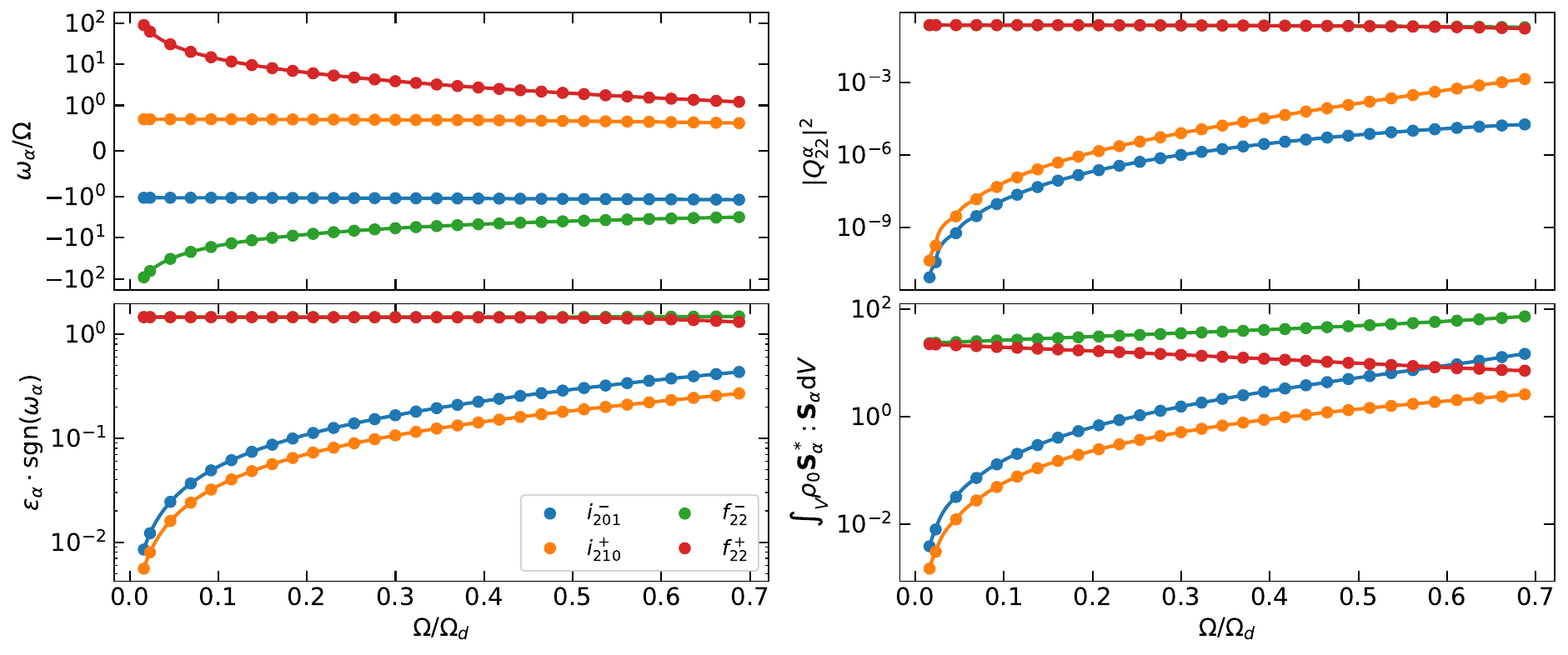}
    \caption{Rotating frame frequencies $\omega_\alpha$ (top left), tidal overlap integrals $Q_{\ell m}^\alpha$ (top right), energy coefficients $\epsilon_\alpha$ (bottom left), and shear tensor contraction integrals (bottom right) computed for the $m=2$ sectoral f-modes (labeled $f_{22}^\pm$, with $+$/$-$ indicating prograde/retrograde), and longest-wavelength $m=2$ inertial modes (labeled $i_{210}^+$ and $i_{201}^-$ for the prograde and retrograde mode, respectively), plotted as a function of rotation rate for the $n=1.5$ polytropic sequence. Note that the ratio $\omega_\alpha/\Omega$ remains roughly constant at all rotation rates for the inertial oscillations: $\omega_\alpha=\beta_\alpha\Omega$, with $\beta_+\approx0.69$ for the prograde mode $i_{210}^+$, and $\beta_-\approx-1.07$ for the retrograde mode $i_{201}^-$.}
    \label{fig:mode}
\end{figure*}

\autoref{fig:mode} plots the mode properties required to compute the dissipative coefficients $\kappa_{22k}$ (see \autoref{eq:kapp}), calculated \citep[as described in ][]{Dewberry2022b} as a function of rotation rate for the polytropic model sequence described in Section \ref{sec:model} and \autoref{app:model}. In particular, $\omega_\alpha$ is the mode frequency in the corotating frame, and
\begin{align}
    \epsilon_\alpha 
    &=\frac{1}{MR^2\Omega_d}
    \left(
    \omega_\alpha
    \langle 
        \boldsymbol{\xi}_\alpha,\boldsymbol{\xi}_{\alpha}
    \rangle
    +\langle 
        \boldsymbol{\xi}_\alpha,
        \text{i}{\bf \Omega\times}\boldsymbol{\xi}_{\alpha}
    \rangle
    \right),
\\
    Q_{\ell m}^\alpha 
    &=\frac{1}{M}
    \langle 
        \boldsymbol{\xi}_\alpha,\nabla[(r/R)^\ell Y_{\ell m}]
    \rangle,
\\ 
    {\bf S}_\alpha 
    &=\frac{1}{2\Omega_d}\left[
        \nabla{\bf v}_\alpha
        +(\nabla{\bf v}_\alpha)^T
        -\frac{2}{3}(\nabla\cdot{\bf v}_\alpha){\bf I}
    \right],
\end{align}
where $\boldsymbol{\xi}_\alpha$ is the Lagrangian displacement associated with a given mode $\alpha$, ${\bf v}_\alpha$ is its velocity perturbation, and 
$\langle \boldsymbol{\xi}_\alpha,\boldsymbol{\xi}_\beta\rangle
=\int_V\rho_0\boldsymbol{\xi}_\alpha^*\cdot\boldsymbol{\xi}_\beta\text{d}V$ defines an inner product. The modes of rotating stars are not orthogonal under this inner product, but we normalize each so that $\langle\boldsymbol{\xi}_\alpha,\boldsymbol{\xi}_\alpha\rangle
=MR^2$.

\begin{figure*}
    \centering
    \includegraphics[width=\textwidth]{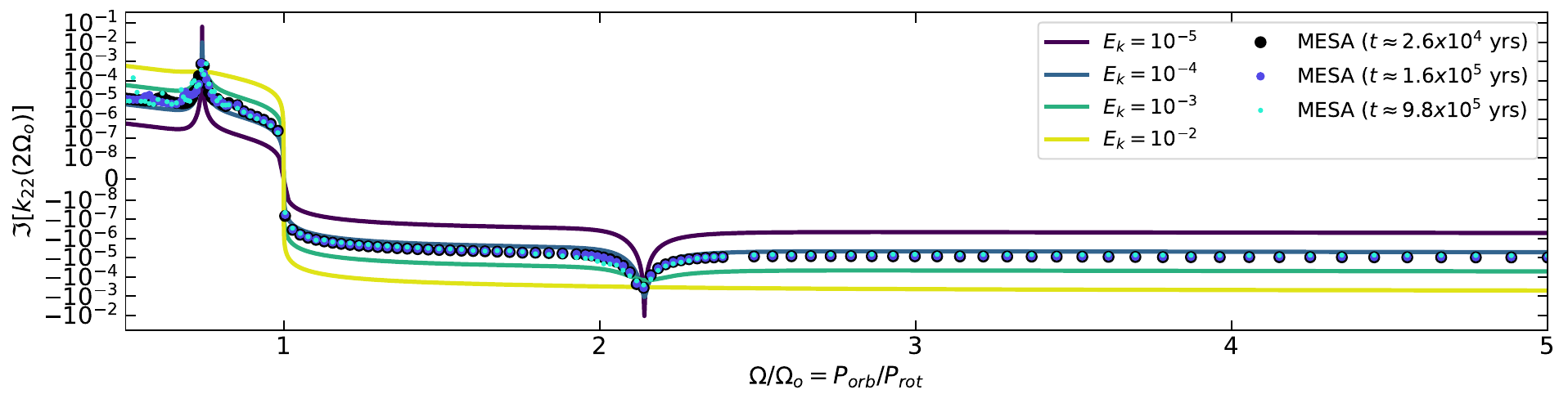}
    \caption{Imaginary parts of tidal Love numbers $\Im[k_{22}(2\Omega_o)]=\kappa_{222}$, plotted as a function of $\Omega/\Omega_o$ for stellar models with $\Omega=0.1\Omega_d$ (the y-axis switches from log to linear scale at $10^{-8}$). The filled points show dissipation coefficients computed \citep[using the approach described in ][]{Dewberry2023} from snapshots of a $1M_\odot$ MESA model, assuming a rotation rate of $\Omega/\Omega_d=0.1$ and ignoring centrifugal flattening, and adopting a kinematic viscosity determined by the convective velocities and mixing lengths. The solid lines show the results of using mode expansions, with mode properties interpolated from those shown in \autoref{fig:mode}, and with different values of Ekman number $E_k=\nu/(R^2\Omega)$ indicated by different line colors.
    }
    \label{fig:k22vs}
\end{figure*}

\autoref{fig:k22vs} plots frequency-dependent profiles of $\Im[k_{22}](2\Omega_o)=\kappa_{222}$ computed using these mode properties and a rotation rate of {$\Omega=0.1\Omega_d$,} for different values of Ekman number (indicated by the colormap). These profiles can be compared with direct tidal dissipation rates computed \citep[using the method described in][]{Dewberry2023} from snapshots of the pre-main sequence MESA model described in \autoref{app:model} (filled points). For the latter calculations, we adopt a spatially variable kinematic viscosity $\nu=\ell_cv_c$, where $\ell_c$ and $v_c$ are the convective mixing length and velocity calculated by MESA. In order to remove discontinuities in the derivative $\partial\nu/\partial r$, we smooth the transition of $\nu$ to zero inside the radiative core with an exponential decay function. 

\autoref{fig:k22vs} demonstrates that our reduced polytropic model captures the essential frequency dependence exhibited by the tidal response computed directly from MESA models on the pre-main sequence. The curves for $E_k=10^{-4}-10^{-3}$ match the direct calculations most closely. We adopt a more conservative $E_k=10^{-5}$ for most of our calculations due to the expectation that short tidal periods should diminish the efficiency of damping by convective turbulence \citep{Zahn1989,Goodman1997,Duguid2020}, and because of uncertainties surrounding the interaction of tidal and convective flows \citep{Terquem2021,Barker2021,Terquem2023}.

\begin{figure*}
    \centering
    \includegraphics[width=\textwidth]{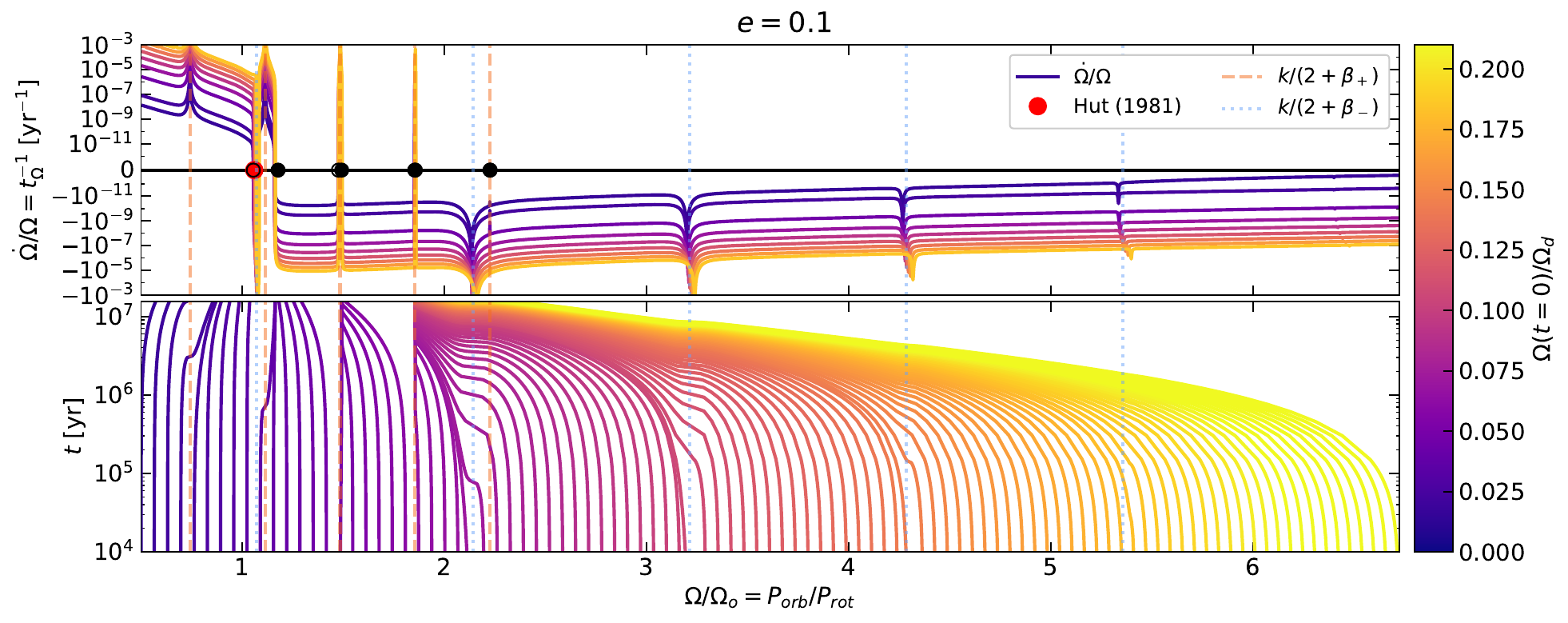}
    \includegraphics[width=\textwidth]{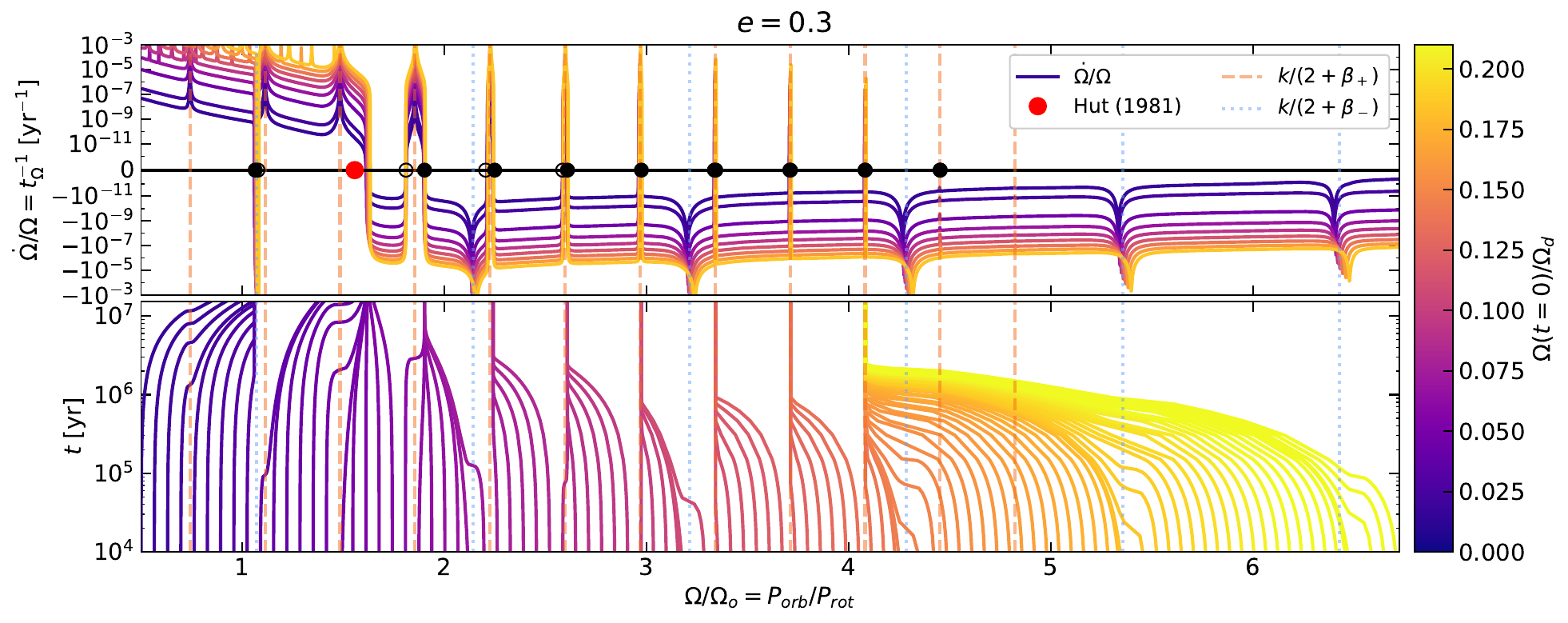}
    \caption{Same as \autoref{fig:e0.5}, but for eccentricities $e=0.1$ (top panels) and $e=0.3$ (bottom panels).}
    \label{fig:evar}
\end{figure*}

\begin{figure*}
    \centering
    \includegraphics[width=\textwidth]{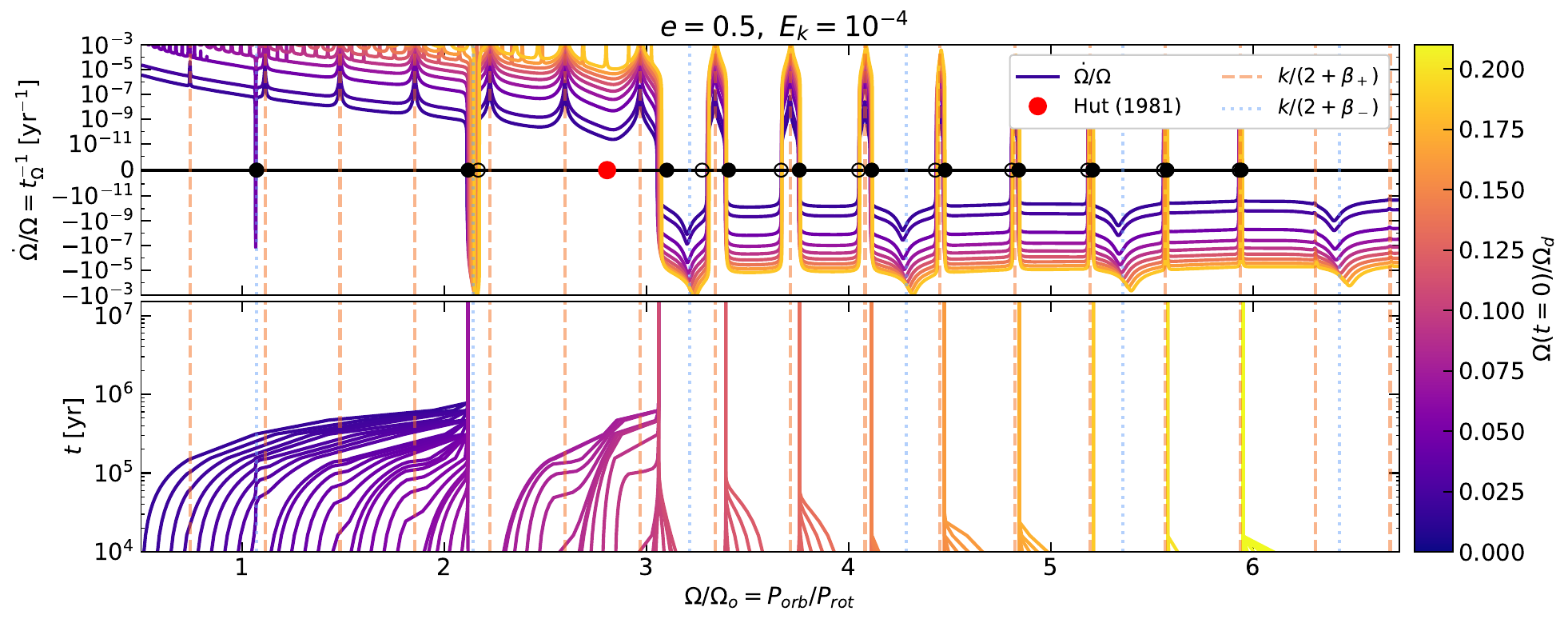}
    \includegraphics[width=\textwidth]{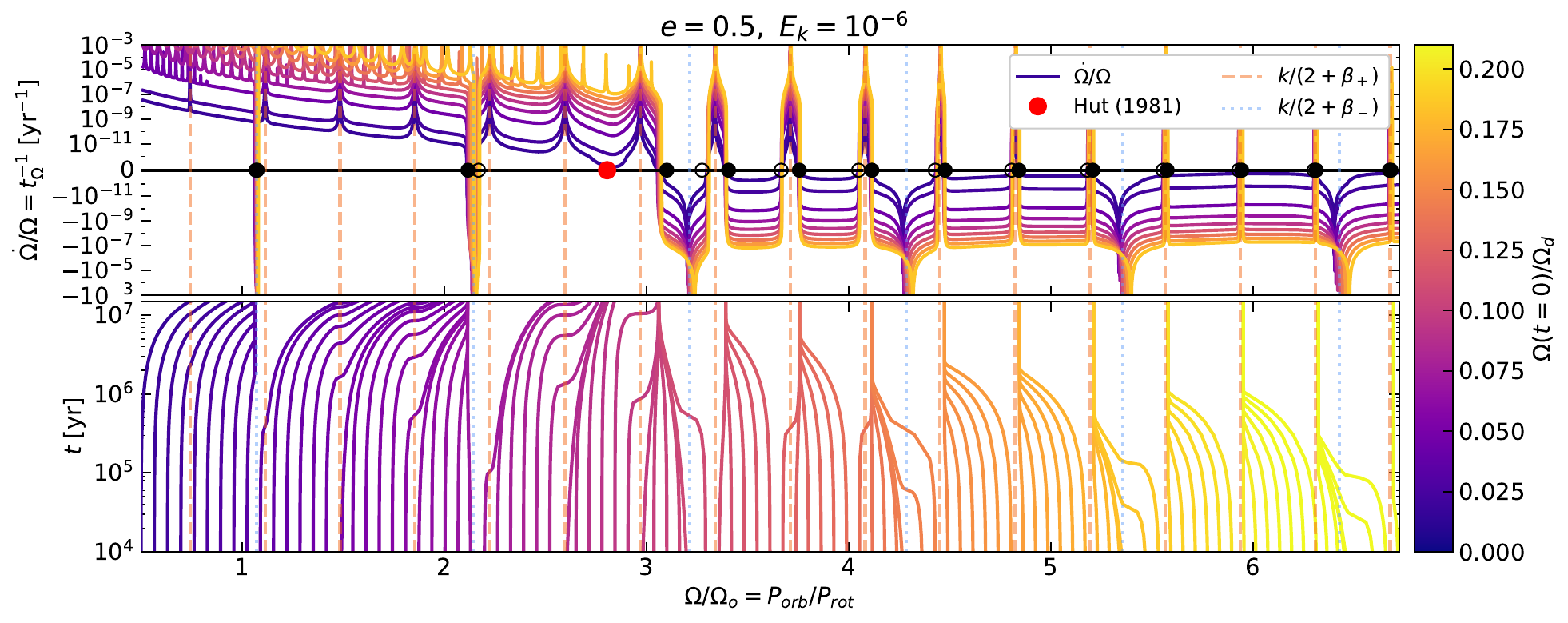}
    \caption{Same as \autoref{fig:e0.5}, but for Ekman numbers $E_k=10^{-4}$ (top panels) and $E_k=10^{-6}$ (bottom panels). For $E_k=10^{-4}$, the $k=1$ synchronization trap for the retrograde inertial mode is overpowered by resonances with the prograde inertial mode for $\Omega/\Omega_d\gtrsim0.05$. Otherwise, the synchronization traps operate as in the case of $E_k=10^{-5}.$}
    \label{fig:Ekvar}
\end{figure*}
\begin{figure*}
    \centering
    \includegraphics[width=0.926\textwidth]{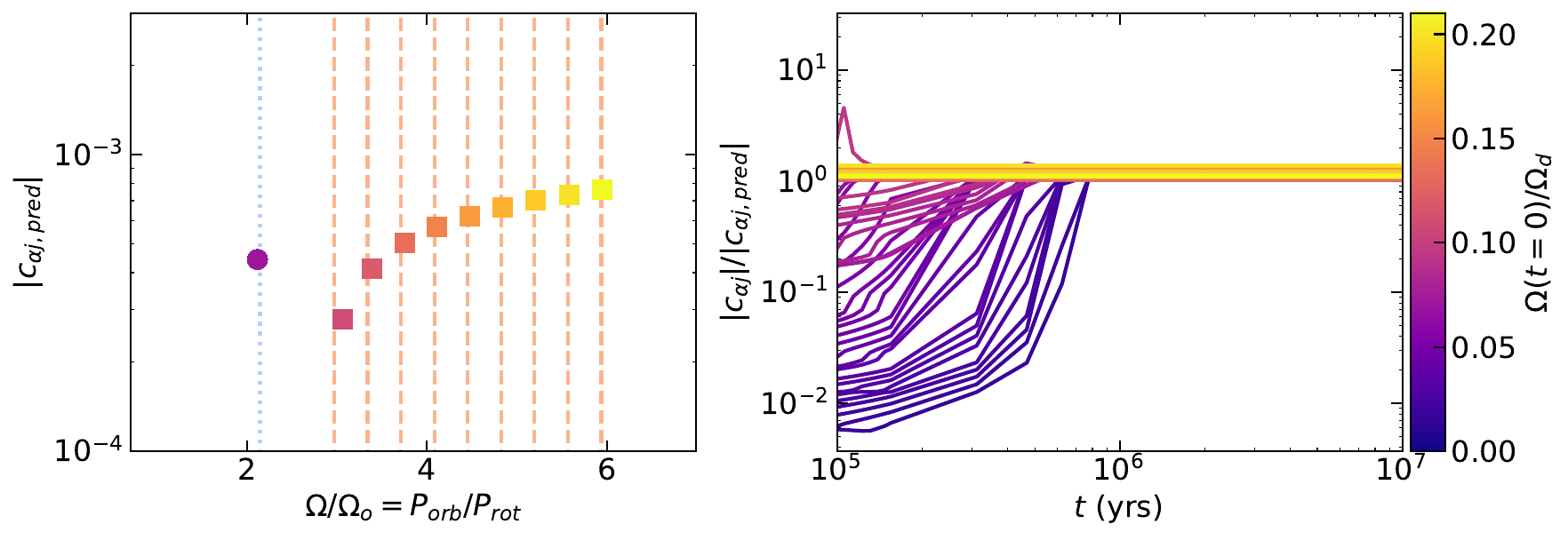}
    \caption{Same as \autoref{fig:amp}, but for Ekman number $E_k=10^{-4}$. Aside from the missing $k=1$ retrograde inertial mode trap (see \autoref{fig:Ekvar}, top), the end-state amplitudes are virtually identical to those computed for $E_k=10^{-5}$.}
    \label{fig:Ek4ca}
\end{figure*}
\section{Varying eccentricity}\label{app:evar}
The panels in \autoref{fig:evar} show the same torques and spin evolution as \autoref{fig:e0.5}, but computed for a $10$ day orbit with eccentricities $e=0.1$ (top) and $0.3$ (bottom). For these lower eccentricities, smaller values of high-$k$ Hansen coefficients produce fewer spin equilibria at which aliased resonances with the inertial modes balance the nonresonant f-mode torque. However, the synchronization traps that do remain still act to inhibit pseudosynchronization. 

\section{Varying viscosity}\label{app:visc}
The panels in \autoref{fig:Ekvar} also replicate those in \autoref{fig:e0.5}, but for Ekman numbers $E_k=10^{-4}$ (top) and $10^{-6}$ (bottom). Fig. \autoref{fig:Ek4ca} plots the same amplitude calculations as \autoref{fig:amp}, but for $E_k=10^{-4}$. The differences in spin evolution primarily come from larger and smaller values for the nonresonant f-mode torques: stronger (weaker) equilibrium tides drive more (less) rapid evolution toward the synchronization traps produced by the inertial oscillations. A qualitative change results from viscosities so large that inertial mode resonances are erased (see the lightest curve in \autoref{fig:k22vs}, and the synchronization trap near $\Omega/\Omega_o\sim1$ in \autoref{fig:Ekvar}, top). %Such strongly damped inertial waves would have little effect on spin synchronization. However, we do not expect the long wavelength inertial modes included in our calculations here to be so strongly damped by convective turbulence. In any case,
However, as demonstrated in \autoref{sec:num} (see Figs. \ref{fig:amp} and \ref{fig:wvar}), for the astrophysically relevant case of weak damping the efficacy of a synchronization trap is effectively independent of the viscosity ascribed to convective turbulence.

\bibliography{sync}{}
\bibliographystyle{aasjournal}



\end{document}